\documentclass[aps, prl, twocolumn, amssymb, amsmath, amsfonts, showpacs, superscriptaddress, preprintnumbers, a4paper]{revtex4-2} 
\usepackage{amssymb, amsmath, amsthm}
\usepackage{xcolor}
\usepackage{graphicx} 
\usepackage{hyperref}
\usepackage{comment}
\usepackage{physics}
\usepackage{footnote}

\begin{document}

\title{Dirac Impurity in a Luttinger liquid}

\author{Lorenzo Gotta}
\email{lorenzo.gotta@unige.ch}
\affiliation{Department of Quantum Matter Physics, University of Geneva, 24 Quai Ernest-Ansermet, 1211 Geneva, Switzerland}

\author{Thierry Giamarchi}
\affiliation{Department of Quantum Matter Physics, University of Geneva, 24 Quai Ernest-Ansermet, 1211 Geneva, Switzerland}

\date{\today}

\begin{abstract} 
We consider a linearly-dispersing quantum impurity interacting through a contact density-density term with a one-dimensional (1D) superfluid described 
by the Tomonaga-Luttinger liquid theory. Using a linked cluster expansion we characterize the impurity dynamics by calculating approximate expressions for the single-particle Green's function and for the time evolution of the density profile. We show the existence of two different dynamical regimes: (i) a \textit{quasiparticle} regime, dominated by the presence of a finite lifetime for the host impurity, and (ii) an \textit{infrared-dominated} regime, where the impurity causes a Anderson's orthogonality catastrophe of the $1D$ bath. We discuss the possible experimental consequences of these findings for cold atoms experiments. 
\end{abstract}

\maketitle

An outstanding question of many-body physics is the characterization of impurity dynamics inside quantum phases of matter~\cite{balatsky04_impurities_review,schmidt_impurities_review,massignan_polarons_review}. The most notable description of the behavior of a distinguishable impurity in a quantum many-body system is the quasiparticle (QP) paradigm, predicting that the propagation of the impurity preserves its coherence over a finite lifetime and the mass of the impurity is renormalized. Such a model has been successful in understanding several phenomena, ranging from the polaron problem \cite{Froehlich_polaron,feynman55_polaron}, that can be realized in cold-atom systems by coupling an impurity atom to the Bogoliubov excitations of a Bose-Einstein condensate~\cite{Bruderer_polarons_optical_lattices,Tempere_BEC_polaron}, and the description of low-energy excitations in interacting electron gases captured by Fermi liquid theory \cite{Coleman2015}. 

However, several analyses have addressed the possibility of a breakdown of QP behavior in the realm of 1D many-body quantum systems \cite{Kopp:1990kn,Rosch_heavy_impurity,zvonarev_ferrobosons07,kamenev_exponents_impurity,kantian_impurity_DMRG,pronko_impurity_fermion,Kamar_impurity_two_leg_ladder}, thus questioning the naive expectation that QP dynamics occurs in all dimensions. In particular, a free impurity interacting with a 1D bath was predicted to propagate subdiffusively \cite{zvonarev_ferrobosons07} within the system, and the mechanism behind such phenomenon was identified with a manifestation of the \emph{Anderson's orthogonality catastrophe} (AOC)~\cite{Anderson1967,Castella_Anderson_1D}, i.e., a reordering of the bath through the impurity-induced generation of a divergent number of low-energy excitations. Since experimental evidence on impurity dynamics in 1D has been accumulating ~\cite{palzer_impurity_1d,catani_impurity_oscillations,fukuhara_heinseberg_cold,fukuhara_magnon_bound_states}, a thorough analytical understanding of possible novel dynamical behavior in such settings 
as well as the observables, such as the density profiles, directly reachable in experiments is required. Indeed contrarily to the single particle correlation function such quantities were mostly 
probed numerically. In addition, in all the previous analyses the quadratic dispersion of the impurity, due to the empty impurity band, was instrumental in the physical 
properties. On the other hand one can expect that at finite filling of the impurity band
one recovers both a linear spectrum and negative energy states, and thus the more conventional Tomonaga-Luttinger liquid (TLL)
features \cite{giamarchi_book_1d} for both species. This makes it particularly interesting to explore the physics of an impurity that would have a linear dispersion relation $\epsilon_i (k)= v_i |k|$, but without neagtive energy states (single particle). Such linear spectra are also directly relevant in a category of materials such as, e.g., graphene.  

In this work, we study the problem of a quantum impurity with such a linear dispersion relation interacting with a 1D quantum system described by a TLL Hamiltonian via a density-density term. We compute both the single particle correlation
function and the density profile of the impurity as a function of time. We show that the linear dispersion relation of 
the impurity leads to novel physics compared to the quadratic one. In particular, although AOC is still 
present when the velocity of the impurity is lower than the one of the bath, it does not lead anymore to a subdiffusive behavior. In the opposite regime, a conventional QP regime is recovered. 

We consider the model Hamiltonian $\hat H = \hat H_{imp}+\hat H_{TLL}+\hat H_{int}$, comprising (i) the Hamiltonian of the bosonic impurity:
$\hat H_{imp} = \sum_k \epsilon_k \hat d^{\dag}_k \hat d_k,$
where the impurity dispersion relation satisfies $\epsilon_{-k}=\epsilon_k$, (ii) the Hamiltonian of the bath, taken to be the one of a TLL \cite{giamarchi_book_1d}:
\begin{equation}
\hat H_{TLL}=\frac{v_s}{2\pi}\int dx \left[K(\partial_x\hat \theta)^2+\frac{1}{K}\left(\partial_x\hat\phi\right)^2 \right],
\end{equation}
where the fields $\hat\phi(x)$, $\hat \theta(x)$ satisfy the commutation relation $[\hat \phi(x),\partial_{x'}\hat \theta(x')]=i\pi\delta(x-x')$, and (iii) the density-density interaction term between the TLL and the bosonic impurity:
\begin{align}
\hat H_{int}= -\frac{g}{\pi} \int dx\,\, (\partial_x \hat \phi)\,\, \hat d^{\dag}(x) \hat d(x),
\end{align}
having introduced $\hat d(x)=\frac{1}{\sqrt{L}}\sum_k e^{ikx} \hat d_k$.

In order to quantify the propagation properties of the impurity inside the $1$D quantum bath, we introduce the notation $\ket{0}$ to denote the vacuum state for both the impurity and the TLL sound modes and evaluate the single-particle Green's function:
\begin{equation} \label{Eq:GF}
 G(y,t;x,0)=\bra{0} \hat d(y) e^{-i\hat H t} \hat d^{\dag}(x) \ket{0},
\end{equation}
which quantifies how drastically the bath gets affected by the propagation of the impurity and the density profile:
\begin{align}
    n(x,t)=\bra{0}\hat d(0) e^{i\hat H t}\hat d^{\dag}(x) \hat d(x) e^{-i\hat H t}\hat d^{\dag}(0)\ket{0}.
\end{align}

A convenient framework for the calculation of time-dependent correlation functions is the Keldysh formalism~\cite{Kamenev2011}, where the object of interest is the Keldysh action $S[\psi,\psi^*]$, appearing in the Keldysh partition function $Z=Tr[\hat\rho(t)]=\int \mathcal{D}[\psi,\psi^*] e^{iS[\psi,\psi^*]}$ and dependent on the retarded and advanced fields $\psi=(\phi_{q,\pm},\varphi_{k,\pm})$ associated to the degrees of freedom in the problem.
The action consists of $3$ terms, identified as the impurity action $S_{imp}$, the action of the TLL bath $S_{TLL}$ and the action $S_{int}$ describing the interaction between the latter two. In the following, we introduce the so called classical and quantum component fields $(\psi_{cl},\psi_{qu})$, respectively obtained from the retarded and advanced fields $\psi_{\pm}$ with identical set of quantum numbers via the rotation $\psi_{cl}=(\psi_+ +\psi_-)/\sqrt{2}$, $\psi_{qu} = (\psi_+-\psi_-)/\sqrt{2}$, where quantum number labels have been omitted for simplicity.

The Keldysh action for the system of noninteracting bosonic modes modelling the bath reads:
\begin{align}\label{Eq:S_TLL}
&S_{TLL}=\sum_{q\neq 0} \int \frac{d\omega}{2\pi} \sum_{\alpha,\beta = cl, qu} \phi^*_{q,\alpha}(\omega)\mathcal{G}^{-1}_{\alpha,\beta}\phi_{q,\beta}(\omega),\\
&\mathcal{G}^{-1}=\begin{pmatrix}
0 &  \omega-\omega_q-i\delta\\
 \omega-\omega_q+i\delta& 2i\delta F(\omega)
\end{pmatrix}.
\end{align}
$F(\omega)=1+2n_B(\omega)$ is simply related to the Bose-Einstein distribution $n_{B}(\omega)$ and $\omega_q=v_s|q|$.
The Keldysh action for the bosonic impurity has a sîmilar structure:
\begin{align}\label{Eq:S_imp}
&S_{imp}=\sum_{k} \int \frac{d\epsilon}{2\pi} \sum_{\alpha,\beta = cl, qu} \varphi^*_{k,\alpha}(\epsilon)\mathcal{G}^{-1}_{\alpha,\beta}\varphi_{k,\beta}(\epsilon),\\
&\mathcal{G}^{-1}=\begin{pmatrix}
0 &  \epsilon-\epsilon_k-i\delta\\
 \epsilon-\epsilon_k+i\delta& 2i\delta F(\epsilon)
\end{pmatrix}.
\end{align}
$F(\epsilon)=1$ when the initial density matrix is the vacuum state $\ket{0}\bra{0}$ for the impurity modes.
Finally, the interaction term takes the form:
\begin{equation}\label{Eq:S_int}
\begin{split}
S_{int} &=\int dt\,\, \left[-H_{int}\left(\phi_{+}(t),\varphi_+(t)\right)+H_{int}\left(\phi_{-}(t),\varphi_-(t)\right) \right] \\
&=-\frac{g\sqrt{K}}{L}\sum_k \sum_{p\neq 0}V(p)\int_{-\infty}^{+\infty} dt\\
&\Biggl\{\varphi^*_{k,+}(t) \varphi_{k+p,+}(t)\left[\phi^*_{p,+}(t)+\phi_{-p,+}(t) \right]+\\
&-\varphi^*_{k,-}(t) \varphi_{k+p,-}(t)\left[\phi^*_{p,-}(t)+\phi_{-p,-}(t) \right] \Biggr\}, 
\end{split}
\end{equation}
where $V(p)=\sqrt{\frac{L|p|}{2\pi}}e^{-\alpha\frac{|p|}{2}}$.

In order to compute the single-particle Green's function \eqref{Eq:GF}, we reexpress it in the Keldysh formalism as:
\begin{multline}
\bra{0} \hat d(y) e^{-i\hat H t} \hat d^{\dag}(x) \ket{0} = \\
\frac{1}{L}\sum_{k_1,k_2} e^{i(k_2 y -k_1 x)}\langle \varphi_{k_2,+}(t) \varphi^{*}_{k_1,+}(0) \rangle,
\end{multline}
where the average is taken over the action obtained by summing \eqref{Eq:S_TLL}, \eqref{Eq:S_imp}, \eqref{Eq:S_int}, together with the vacuum conditions $F(\omega)=1$ in the expression of $S_{TLL}$ and $F(\epsilon)=1$ in $S_{imp}$. A second-order perturbative expansion in the interaction $S_{int}$ followed by a reexponentiation of the result, in the spirit of the linked-cluster expansion \cite{mahan_book}, leads to the final expression:
\begin{equation}\label{Eq:GF_LCE}
    G(y,t;x,0)\approx \int \frac{dk}{2\pi}e^{ik(y-x)-i\epsilon_k t-\int du R(u;k)\frac{1-iut-e^{-iut}}{u^2}},
\end{equation}
where $R(u;k)=\frac{g^2 K}{(2\pi)^2}\int dp |p| e^{-\alpha|p|}\delta(u-\epsilon_{k+p}-\omega_p+\epsilon_k)$ (see SM~\ref{SM:sec:GF_LCE}).

Although the previous expressions were generic, we specialize from now on to the case of a linear dispersion relation for the impurity, namely $\epsilon_i (k)=v_i |k|$. In such case, the interaction-dependent term in the exponent of the integrand in \eqref{Eq:GF_LCE} becomes easily treatable and allows for the estimate of the long-time behavior of the momentum-space Green's function (see SM \ref{SM:sec:GF_k}):
\begin{align} \label{Eq:MSGF}
    G(k,t;k,0)\approx e^{-iv_i |k| t-\int du R(u;k)\frac{1-iut-e^{-iut}}{u^2}},
\end{align}
that can be read off from \eqref{Eq:GF_LCE}. 

One can distinguish two regimes, depending on whether $v_i>v_s$ or $v_i<v_s$. The numerical evaluation of $|G(k,t;k,0)|$ in the two regimes shows that the asymptotic behavior of the momentum-space Green's function is dramatically different. This can be traced back to the fact that the impurity energy difference $\delta\epsilon_{k,q}:=v_i|k|-v_i |k+q|$ intersects the phonon dispersion relation $\omega_q=v_s |q|$ only at $q=0$ when $v_i<v_s$, thus only causing the quasiresonant generation of vanishingly-small-momentum phonons. On the other hand, when $v_i>v_s$, one observes that $\delta\epsilon_{k,q}=v_s |q|$ also for a finite nonzero value $q^*=-\frac{2v_i}{v_i+v_s}k$, thus signaling that the impurity emits resonantly finite-momentum phonons as well in this regime.
(\ref{Eq:MSGF}) also reveals (see (\ref{SM:Eq:GF_QP})) two distinct possible behaviors: when $v_i>v_s$, the large-momentum and large-time behavior of $|G(k,t;k,0)|$ decays exponentially as:
\begin{multline}\label{Eq:GF_QP}
  |G(k,t;k,0)|\sim \\
  e^{-Re\{F(t)\}}e^{\frac{g^2 K}{(2\pi)^2} \int_k^{+\infty} dq\,  \frac{e^{-\alpha q}}{(v_s-v_i)^2 q}}e^{-\frac{g^2 K}{(2\pi)^2}\frac{2\pi v_i k e^{-\frac{2\alpha v_i k}{v_i+v_s}}}{(v_i+v_s)^2}t}, 
\end{multline}
where the expression of $F(t)$ is reported in \eqref{SM:Eq:F_t}.
The presence of a finite-momentum-phonon emission channel in this regime is quantitatively signaled by the quasiparticle decay term of the form $e^{-\frac{t}{\tau(k)}}$, where the expression of the lifetime $\tau(k)$  can be read off from the exponent of the last term of \eqref{Eq:GF_QP}. This reveals that $\tau(k)$ decreases with interaction strength as $g^{-2}$ and, in the limit of vanishing momentum cutoff $\alpha$, with the impurity momentum $k^{-1}$. Moreover, when the quasiparticle resonant momentum $q^*=-\frac{2v_i k}{v_i+v_s}$ (i.e., the nonzero momentum satisfying $\delta\epsilon_{k,q^*}=v_i|q^*|$) satisfies $|q^*|\gg \alpha^{-1}$, the momentum exchange between the $1$D bath and the impurity is strongly suppressed due to the existence of a finite momentum cutoff on the bath excitation modes, thus recovering, within an extended transient regime, the dynamical properties that the impurity displays  when the relation $\delta\epsilon_{k,q}=v_s|q|$ is satisfied only by $q=0$ for all values of $k$, to be described in the following paragraph.

Contrarily, when $v_i<v_s$, the absence of a nonzero momentum satisfying the scattering resonance condition results in a different behavior of $|G(k,t;k,0)|$ at large time; when $t\gg 1$, one has (see \eqref{SM:Eq:GF_IR}):
\begin{align} \label{Eq:GF_IR}
    &|G(k,t;k,0)|\sim \nonumber\\
    &\sim e^{-Re\{F(t)\}}e^{-\frac{g^2 K}{(2\pi)^2}\int_k^{+\infty}dq \, q e^{-\alpha q} \left(\frac{1}{[(v_i+v_s)q-2v_i k]^2}-\frac{1}{(v_s-v_i)^2 q^2}\right)},
\end{align}
which reproduces a slow power-law decay in time associated to the generation of phonons with vanishingly small momentum. The latter grow logarithmically (rather than linearly) in time as a consequence of the vanishing number of low-momentum particle-hole scattering processes on top of a filled Fermi sea in linearized bands described by TLL theory.

\begin{figure}
\includegraphics[width=0.5\textwidth]{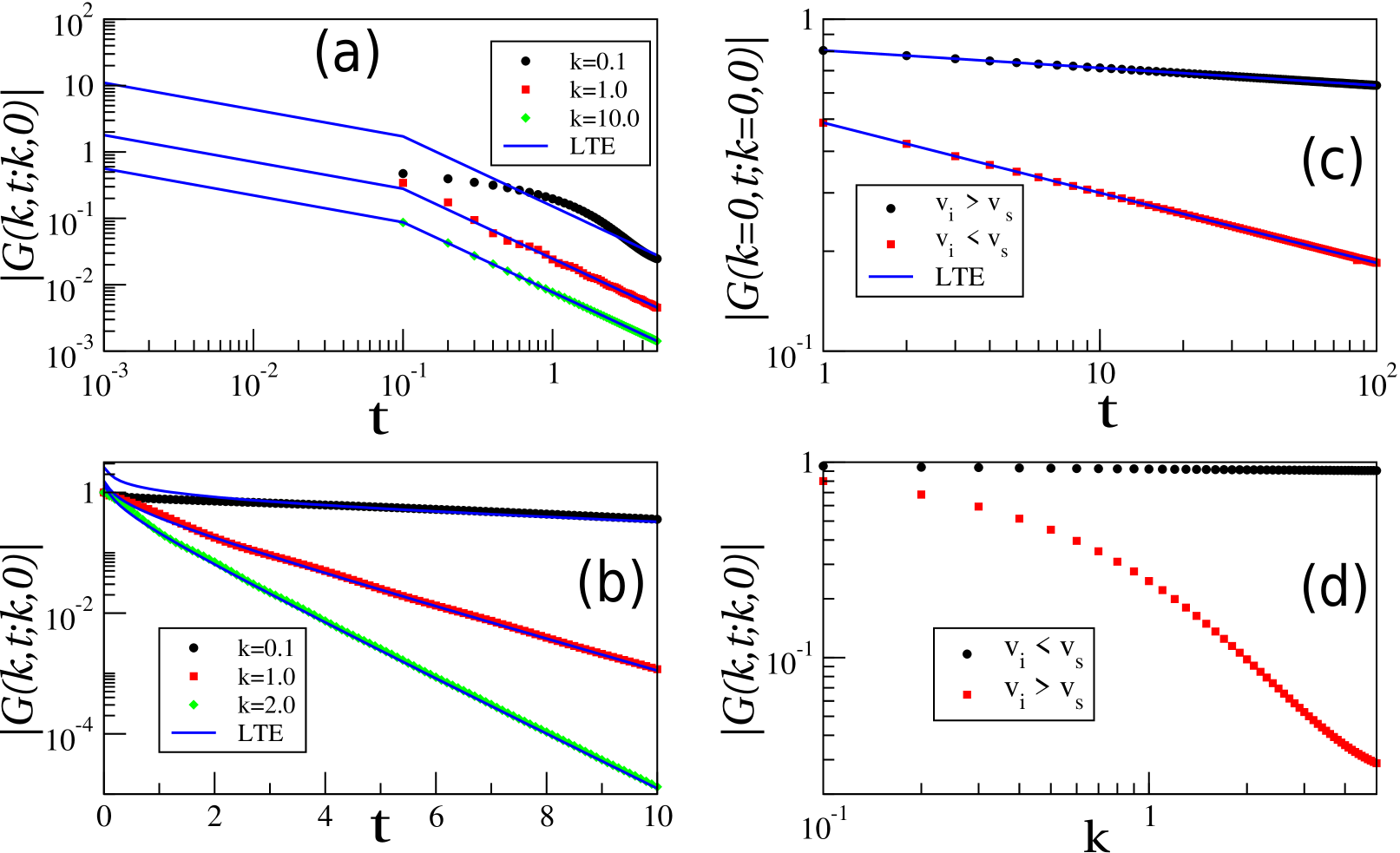}
\caption{(a)-(b)-(c): $|G(k,t;k,0)|$ vs $t$ for (a) $v_i<v_s,\, k>0$, (b) $v_i>v_s,\, k>0$ and (c) $v_i\lessgtr v_s,\, k=0$ compared to and in agreement with their large-time analytical estimates in \eqref{Eq:GF_QP},\eqref{Eq:GF_IR} and \eqref{Eq:GF_0}. We use $v_s=2,\, g=5.0,\, K=1.5,\,\alpha=0.1$ and (a) $ v_i=v_s/2$, (b) $ v_i=2v_s$  and (c) $ v_i=v_s/2$ and $ v_i=2v_s$. (d): $|G(k,t;k,0)|$ vs $k$ for $t=10.0,\, v_i=4.0,\, g=2.0,\, \alpha=0.1,\, K=1.5$, with $v_s=1.0,\, 7.0$.}
\label{Fig:MSGF}
\end{figure}

We plot in Fig.~\ref{Fig:MSGF}(up) the curve $|G(k,t;k,0)|$ as a function of $t$ for distinct values of $k$ and different choices in the relative magnitude of $v_i$ and $v_s$. 
We confirm the analytical predictions of (\ref{Eq:GF_QP}-\ref{Eq:GF_IR}) by comparing them to the direct numerical evaluation of $|G(k,t;k,0)|$ in \eqref{Eq:MSGF}. For sufficiently large time $t$ we obtain curves that reproduce an exponential time-decay of $|G(k,t;k,0)|$ within the $v_i>v_s$ regime and a slow power-law decay within the $v_i<v_s$ regime. Moreover, we compare in Fig.~\ref{Fig:MSGF}(down) the behavior of $|G(k,t;k,0)|$ as a function of $k$ for a fixed finite value of $t$ in the two identified dynamical regimes. We manifestly see the significantly faster decrease of the finite-$k$ values of the momentum-space Green's function in the $v_i>v_s$ regime of motion, as compared to the slower decay achieved in the $v_i<v_s$ case. 

The zero-momentum behavior of $|G(k,t;k,0)|$, is directly linked to the generation of low-energy phonons by an impurity with $k=0$, since the latter, regardless of the relative magnitude of $v_i$ and $v_s$, meets the resonance condition $\delta\epsilon_{k=0,q}=v_s|q|$ only at $q=0$. A careful analysis of the expression in \eqref{Eq:MSGF} at $k=0$ leads to the large-time estimate:
\begin{equation}\label{Eq:GF_0}
    |G(k=0,t;k=0,0)|\sim e^{-\frac{g^2 K}{(2\pi)^2}\frac{2}{(v_i+v_s)^2}\log\left(\frac{(v_i+v_s)t}{\alpha}\right)},
\end{equation}
namely, a power law decay in time (see SM~\eqref{SM:sec:GF_0}). We confirm this result by comparing the numerical evaluation of \eqref{Eq:MSGF} with $k=0$ to the analytical estimate in \eqref{Eq:GF_0}. The result is shown in Fig.~\ref{Fig:MSGF}, for $v_i>v_s$ and $v_i<v_s$. 

\begin{figure}
\includegraphics[width=0.5\textwidth]{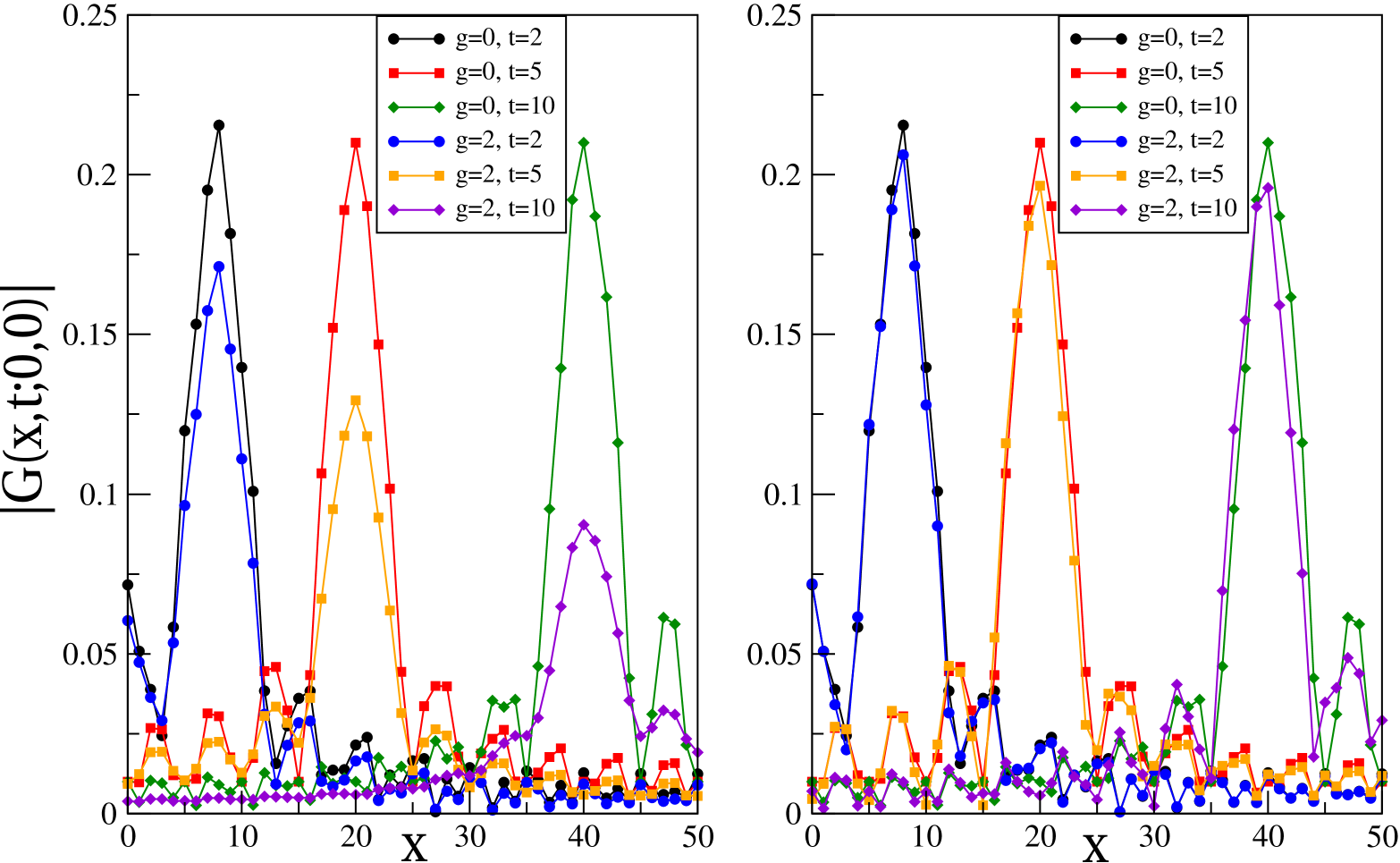}
\caption{Left: time evolution of $|G(x,t;0,0)|$ as a function of time $t$ in the case $v_i>v_s$ for parameter values $L=100,\, v_s=1,\, v_i=4,\, K=1.5,\, \alpha=10^{-1}$, both for $g=0$ and $g=2$. Right: time evolution of $|G(x,t;0,0)|$ as a function of time $t$ in the case $v_i<v_s$ for parameter values $v_s=7$, the other parameters remaining unchanged, both for $g=0$ and $g=2$.}
\label{Fig:GF_t}
\end{figure}

Let us now focus on the real-space behavior of the Green's function. We fix the values of the parameters of the problem with the sole exception of the phonon velocity $v_s$, which satisfies alternatively $v_i>v_s$ or $v_i<v_s$. We plot the result in Fig.~\ref{Fig:GF_t}.
For $v_i>v_s$ a rapid decay of the amplitude of the peak in $|G(y,t;x,0)|$ around $x_p\approx v_i t$ occurs. It is absent when $v_i<v_s$, since the finite-momentum impurity Green's function is exponentially suppressed in the large-time limit when $v_i> v_s$. The impurity coherence is thus more rapidly destroyed by hybridization between the impurity and the bath degrees of freedom. We also show in Fig.~\ref{SM:Fig:GF_g} the effect of an increasing interaction strength on the propagation speed and coherence of the impurity. The interactions both slow down the impurity propagation and destroy its quasiparticle coherence. The effect is much more dramatic in the case $v_i>v_s$, as noticed before.

Let us now turn to the density profiles computed again via the linked cluster 
expansion (see SM~\eqref{SM:density}). Such density profiles can be measured in cold atom experiments~\cite{catani_impurity_oscillations,fukuhara_heinseberg_cold,fukuhara_magnon_bound_states,meinert_quench,haller_center_super_bloch,juergensen_correlated_tunneling}.
The resulting expressions are valid for any dispersion relation satisfying $\epsilon_{-k}=\epsilon_k$. The initial condition for the impurity is a wavepacket of amplitude $\psi(y)$ of the form $\ket{\psi_{imp}(0)}=\int_{-L/2}^{L/2} dy \,\psi(y)\, \hat d^{\dag}(y)\ket{0}$, while the $1D$ quantum bath is at zero temperature. The local density expectation value within this approximation reads:
\begin{align}
   &\bra{\psi(t)}\hat d^{\dag}(x) \hat d(x) \ket{\psi(t)}=\nonumber\\
   &\sum_{k,k_2}(\mathcal{F}\psi)^*(k_2)(\mathcal{F}\psi)(k) e^{i[(kx-\epsilon_k t) - (k_2 x-\epsilon_{k_2}t)]}\nonumber\\
& \exp\Biggl\{-\Biggl[\int du R(u;k)\frac{1-iut-e^{-iut}}{u^2}+\nonumber\\
&+\int du R(u;k_2)\frac{1+iut-e^{iut}}{u^2}+\nonumber\\
&-\int du\int dv S(u,v;k,k_2) \frac{1-e^{iut}}{u}\frac{1-e^{-ivt}}{v}\Biggr]\Biggr\}, \label{Eq:density_LCE}
\end{align}
where we introduced the function $S(u,v;k,k_2)=\frac{g^2 K}{(2\pi)^2}\int dp |p| e^{-\alpha|p|}\delta(u-\epsilon_{k_2+p}-\omega_p+\epsilon_{k_2})\delta(v-\epsilon_{k+p}-\omega_p+\epsilon_k)$ and the Fourier transform of the initial wavefunction $(\mathcal{F}\psi)(q)=\frac{1}{L}\int_{-L/2}^{L/2} dx\, e^{-iqx}\psi(x)$ (see SM~\ref{SM:density}). The expression is normalized regardless of the specific form of the dispersion relation. 

As before, we now specialize to the case of a linear dispersion relation $\epsilon_k=v_i |k|$. We assume the Fourier transform of the initial wave-packet to be a gaussian centered at momentum $k_0$ with variance $\sigma^2$, i.e., $(\mathcal{F}\psi)(k)=\frac{(2\pi)^{1/4}}{\sqrt{L^2\sigma}}e^{-(k-k_0)^2/(2\sigma^2)}$. The propagation of the wavepacket is strongly affected by the zeros of the function $v_i|k+q|+v_s|k|-v_i|k|$. We fix the values of the parameters of the problem with the sole exception of the phonon velocity $v_s$, which satisfies alternatively $v_i>v_s$ or $v_i<v_s$. When $v_i>v_s$, the presence of a nonzero solution to the equation $v_i|k+q|+v_s|k|-v_i|k|=0$ is linked to the damping of the peak in the density during time evolution. This behavior can be linked to (\ref{Eq:GF_QP}), that shows that the presence of a nonzero solution to the scattering resonance condition allows for quasiparticle behavior with a finite lifetime. On the other hand, when $v_i<v_s$, the propagation of the density wavepacket is almost unaffected by the impurity-bath interaction. In such case, the impurity can only quasi-resonantly emit low-momentum phonons, which are injected into the bath as in the paradigmatic AOC problem. 

\begin{figure}
\includegraphics[width=0.5\textwidth]{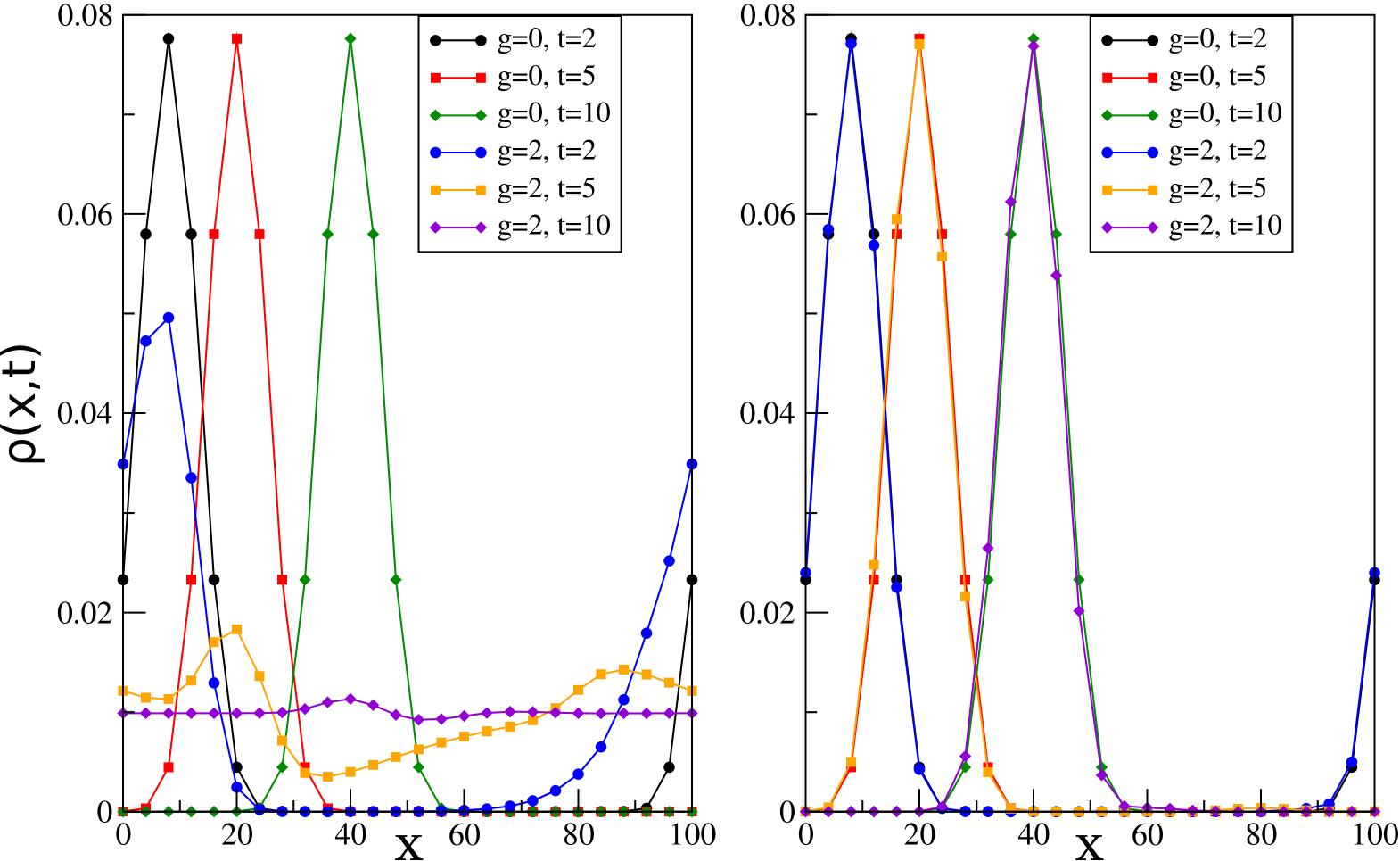}
\caption{\label{Fig:n_t}
Left: evolution of the density profile $\rho(x,t)$ vs $t$ in the case $v_i>v_s$ for parameter values $L=100, k_0=\pi/2,\, v_s=1,\, v_i=4,\, K=1.5,\, \alpha=10^{-1}, \,\sigma=0.1$, both for $g=0$ and $g=2$. Right: same plot as above in the case $v_i<v_s=7$, the other parameter values being unchanged.}
\end{figure}

A comparison of the time evolution of the impurity density profile in absence of impurity-bath interactions to its counterpart in presence of a finite interaction strength $g$, both for $v_i>v_s$ and $v_i<v_s$ is shown in Fig.~\ref{Fig:n_t}. The result is consistent with the data in Fig.~\ref{Fig:GF_t} and shows a strong damping of the ballistically-propagating peak of the initial density profile when $v_i>v_s$, and an almost free evolution when $v_i<v_s$ instead.

Previous work on the single-particle Green's function of a free impurity ($\epsilon_k=t_i k^2$)\cite{kantian_impurity_DMRG} is consistent with our results on $|G(k,t;k,0)|$, as one can infer by defining $v_i(k)=2t_i k$ as the group velocity associated to a quadratic dispersion relation, and comparing it to $v_s$ to identify distinct dynamical regimes. More interestingly, the small-$k$ dispersion of the impurity velocity $v_i(k)$ has dramatic consequences on the asymptotic behavior of $|G(x,t;0,0)|$ for large values of $x$ and $t$ compared to the $\epsilon_k=v_i |k|$ case, as it determines a different low-$k$ behavior of $|G(k,t;k,0)|$. The  subdiffusive spread of correlations for a free impurity is a consequence of the linearly-vanishing group velocity around $k=0$ \cite{kantian_impurity_DMRG}; in the case of a linear dispersion relation, instead, $v_i$ is constant and the $k$-dependence appears only in the form of a term that is $O(1)$ in time, which does not significantly alter the non-interacting case of ballistic propagation. Moreover, we point out that the result in Fig.~\ref{Fig:n_t} for the time evolution of the impurity density point towards a marginal effect of impurity-bath interactions on $\rho(x,t)$ when $v_i<v_s$. Such a signature cannot be directly inferred from the behavior of $G(x,t;0,0)$ and showcases a form of emergent protection of free propagation that is absent in the quasiparticle ($v_i>v_s$) dynamical regime.

This predictions can potentially be tested in cold atomic systems. The main difficulty is to realize a Dirac dispersion for the impurity, but several 
proposals have been already put forward\cite{lu_floquet_dirac_bands,kuno_generalized_dirac_simulation,cirac_cold_atom_qft_dirac}, where the additional challenge of removing negative energy states from the Dirac bands stands. Quantum microscope allow for a direct measurement of the density 
profile and a comparison with the results of Fig,~\ref{Fig:n_t}. The velocity of the bath could be easily varried by changing e.g. the interaction between 
the particles of bath. 

In this work, we have characterized the properties of a mobile impurity with linear dispersion relation inside a 1D quantum system, described by the Luttinger liquid low-energy universality class. By using a linked-cluster expansion approximation, we have computed  the single-particle Green's function and the density profile of the impurity,  This unveils a transition in the dynamical properties of the impurity as a function of the relative magnitude of the impurity velocity $v_i$ and of the phonon sound velocity $v_s$. When $v_i>v_s$, the impurity acquires a finite lifetime due to finite-momentum phonon emission and thus displays the characteristic behavior of a quasiparticle; when $v_i<v_s$, the impurity propagation is roughly unaffected, while the quasi-resonant emission of low-momentum phonons into the bath shakes the low-energy excitation occupations within the bath itself and results in an AOC. 

The current work opens perspective lines of research in the identification of 1D many-body quantum phases of matter through impurity dynamics, as it improves the understanding of the interplay between the impurity dispersion relation and 1D superfuidity. Future works may be concerned with the characterization of impurity dynamics within different phases of matter that can be realized in 1D settings, such as Mott insulators and Bose glasses. Several extensions of our work would be desirable, such as taking into account the finite temperature effects, or extending the calculations of the density
profile to other dispersions. 

\textit{Acknowledgements}: This work was supported in part by the Swiss National Science Foundation under 
grants 200020-188687 and 200020-219400. 
TG would like to thank the program on ``New Directions in far from Equilibrium Integrability and beyond'' and the Simons Center for Geometry and Physics, Stony Brook University, for support and hospitality. 

\bibliography{totphys-A-J,totphys-K-Z,lorenzo}

\newpage
\clearpage
\renewcommand{\thefigure}{S\arabic{figure}}
\renewcommand{\thesection}{S\arabic{section}}
\renewcommand{\theequation}{S\arabic{equation}}
\renewcommand{\thepage}{S\arabic{page}}
\setcounter{page}{0}
\setcounter{section}{0}
\setcounter{secnumdepth}{2}

\onecolumngrid
\begin{center}
{\large \textbf{Online supplementary material for: \\ Dirac impurity in a Luttinger liquid}}

\vspace{16pt}

Lorenzo Gotta$^{1}$, Thierry Giamarchi$^{1}$

\begin{small}
$^1$\textit{ Department of Quantum Matter Physics, University of Geneva, 24 Quai Ernest-Ansermet, 1211 Geneva, Switzerland}
\end{small}

\vspace{10pt}

\today

\vspace{10pt}

\end{center}

\section{Derivation of the Green's function in real space} \label{SM:sec:GF_LCE}

We sketch here the derivation of the Green's function of the impurity shown in \eqref{Eq:GF_LCE}. We construct an estimate of the single-particle Green's function of the impurity in real space:
\begin{equation} \label{Eq:quad_avg}
\bra{0} \hat d(y) e^{-i\hat H t} \hat d^{\dag}(x) \ket{0}=\frac{1}{L}\sum_{k_1,k_2} e^{i(k_2 y -k_1 x)}\bra{0} \hat d_{k_2}(t) \hat d^{\dag}_{k_1}(0) \ket{0}=\frac{1}{L}\sum_{k_1,k_2} e^{i(k_2 y -k_1 x)}\langle \varphi_{k_2,+}(t) \varphi^{*}_{k_1,+}(0) \rangle,
\end{equation}
by computing a second-order perturbative estimate of the average in the last expression in~\eqref{Eq:quad_avg}, and then performing a nonperturbative exponentiation of the result. The average reads:
\begin{align}
\begin{split}
    & \int \mathcal{D}\phi \mathcal{D}\varphi \, e^{i(S_0+S_{int})} \varphi_{k_2,+}(t) \varphi^{*}_{k_1,+}(0) =\langle \varphi_{k_2,+}(t) \varphi^{*}_{k_1,+}(0)  e^{iS_{int}} \rangle_{S_0}=\\
    &=\langle \varphi_{k_2,+}(t) \varphi^{*}_{k_1,+}(0)   \rangle_{S_0}-\frac{1}{2}\langle \varphi_{k_2,+}(t) \varphi^{*}_{k_1,+}(0) S^2_{int}  \rangle_{S_0},
    \end{split}
\end{align}
where the first-order term vanishes because it contains the average of single phonon field terms over the quadratic TLL action. 
We employ the expressions for the impurity field free Green's functions:
\begin{align}
& \langle \varphi_{k,\pm}(t)\varphi^*_{k',\mp}(t')\rangle_{S_0}= \delta_{k,k'}e^{-i\epsilon_k(t-t')}\frac{F(\epsilon_k)\mp 1}{2},\label{SM:Eq:corr1}\\
&\langle  \varphi_{k,\pm}(t)\varphi^*_{k',\pm}(t') \rangle_{S_0} =\theta(t-t')\delta_{k,k'}e^{-i\epsilon_k(t-t')}\frac{F(\epsilon_k)\pm 1}{2}+\theta(t'-t)\delta_{k,k'}e^{-i\epsilon_k(t-t')}\frac{F(\epsilon_k)\mp 1}{2},\label{SM:Eq:corr2}
\end{align}
and similarly for the phonon fields:
\begin{align}
& \langle \phi_{q,\pm}(t)\phi^*_{q',\mp}(t')\rangle_{S_0}= \delta_{q,q'}e^{-i\omega_q(t-t')}\frac{F(\omega_q)\mp 1}{2},\label{SM:Eq:corr3}\\
&\langle  \phi_{q,\pm}(t)\phi^*_{q',\pm}(t') \rangle_{S_0} =\theta(t-t')\delta_{q,q'}e^{-i\omega_q(t-t')}\frac{F(\omega_q)\pm 1}{2}+\theta(t'-t)\delta_{q,q'}e^{-i\omega_q(t-t')}\frac{F(\omega_q)\mp 1}{2},\label{SM:Eq:corr4}
\end{align}
where $F(\epsilon)=1+2n_B (\epsilon)$ and $n_B(\epsilon)=(e^{\beta\epsilon}-1)^{-1}$ is the Bose-Einstein distribution function. Using Wick's theorem, one obtains the result:
\begin{align}
\begin{split}
&\bra{0} \hat d(y) e^{-i\hat H t} \hat d^{\dag}(x) \ket{0}= \frac{1}{L}\sum_{k_1,k_2}e^{i(k_2 y-k_1 x)} \delta_{k_1,k_2} e^{-i\epsilon_{k_1}t} \left[1-\int du R(u;k_1)\frac{1-iut-e^{-iut}}{u^2} \right]\approx\\
&\approx \int \frac{dk}{2\pi}e^{ik(y-x)-i\epsilon_k t-\int du R(u;k)\frac{1-iut-e^{-iut}}{u^2}},\label{SM:Eq:RSGF}
\end{split}
\end{align}
where $R(u;k)$ is defined as:
\begin{equation}
R(u;k)=\frac{g^2 K}{(2\pi)^2}\int dp |p| e^{-\alpha|p|}\delta(u-\epsilon_{k+p}-\omega_p+\epsilon_k),\label{Eq:R}
\end{equation}
and in the second equality we have reexponentiated the term in square brackets, in the spirit of the linked-cluster expansion method.
The momentum-space Green's function is read off from the expression of the real-space Green's function in ~\eqref{SM:Eq:RSGF} as its inverse Fourier transform and takes the form:
\begin{align}
  G(k,t;k,0)\approx e^{-iv_i |k| t-\int du R(u;k)\frac{1-iut-e^{-iut}}{u^2}}.  
\end{align}

\section{Derivation of the Green's function in momentum space} \label{SM:sec:GF_k}

If one chooses the dispersion relation of the impurity to be linear, i.e., $\epsilon_k=v_i |k|$, one can compute the nontrivial term in the exponent in~\eqref{SM:Eq:RSGF} for $k>0$ ( without loss of generality due to the property $R(u;-k)=R(u;k)$) as:
\begin{align}
\begin{split}
&\exp{-\int du R(u;k)\frac{1-iut-e^{-iut}}{u^2}}=\\
& \exp\Biggl\{-\frac{g^2 K}{(2\pi)^2}\Biggl[\int_k^{+\infty} dq\,q e^{-\alpha q}\frac{1-it[(v_i+v_s)q-2v_i k]-e^{-it[(v_i+v_s)q-2v_i k]}}{[(v_i+v_s)q-2v_i k]^2}+\\
& -\int_k^{+\infty} dq\, qe^{-\alpha q}\frac{1-i(v_s-v_i)qt -e^{-i(v_s-v_i)qt}}{(v_s-v_i)^2 q^2} +\frac{1}{(v_i+v_s)^2} \log \left(1+i\frac{v_i+v_s}{\alpha}t\right)+\frac{1}{(v_s-v_i)^2} \log \left(1+i\frac{v_s-v_i}{\alpha}t\right)+\\
&-i\frac{t}{(v_i+v_s)\alpha}-i\frac{t}{(v_s-v_i)\alpha}\Biggr]\Biggr\}.\label{SM:Eq:GFk}
\end{split}
\end{align}

\subsection{The case $v_i>v_s$}

If $v_i>v_s$, then the maximum of the real part of the first integrand of~\eqref{SM:Eq:GFk} $q^*=\frac{2v_i}{v_i+v_s}k$ falls within the domain of integration. Exploiting the fact that:
\begin{align}
    \lim_{t\rightarrow +\infty} \frac{1}{\pi t}\frac{1-\cos(tx)}{ x^2}= \delta(x)
\end{align}
in the first integral in the modulus of~\eqref{SM:Eq:GFk} and neglecting the oscillating term in the second integral of the same expression, one obtains the final result displayed in~\eqref{Eq:GF_QP}:
\begin{align}
  \Biggl|\exp{-\int du R(u;k)\frac{1-iut-e^{-iut}}{u^2}}\Biggr|= \sim e^{-Re\{F(t)\}}e^{\frac{g^2 K}{(2\pi)^2} \int_k^{+\infty} dq\,  \frac{e^{-\alpha q}}{(v_s-v_i)^2 q}}e^{-\frac{g^2 K}{(2\pi)^2}\frac{2\pi v_i k e^{-\frac{2\alpha v_i k}{v_i+v_s}}}{(v_i+v_s)^2}t}, \label{SM:Eq:GF_QP}
\end{align}
where:
\begin{align}
   &F(t)=\frac{g^2 K}{(2\pi)^2}\left[\frac{1}{(v_i+v_s)^2} \log \left(1+i\frac{v_i+v_s}{\alpha}t\right)+\frac{1}{(v_s-v_i)^2} \log \left(1+i\frac{v_s-v_i}{\alpha}t\right)-i\frac{t}{(v_i+v_s)\alpha}-i\frac{t}{(v_s-v_i)\alpha}\right]. \label{SM:Eq:F_t} 
\end{align}

\subsection{The case $v_i<v_s$}

If $v_i<v_s$, instead, the maximum of the real part of the first integrand of ~\eqref{SM:Eq:GFk} $q^*=\frac{2v_i}{v_i+v_s}k$ falls outside of the domain of integration. Then, the oscillating terms in the first and second integrals of ~\eqref{SM:Eq:GFk} may be neglected in the large-time limit and one obtains the r.h.s. of ~\eqref{Eq:GF_IR}:
\begin{align}
   \Biggl|\exp{-\int du R(u;k)\frac{1-iut-e^{-iut}}{u^2}}\Biggr| \sim e^{-Re\{F(t)\}}e^{-\frac{g^2 K}{(2\pi)^2}\int_k^{+\infty}dq \, q e^{-\alpha q} \left(\frac{1}{[(v_i+v_s)q-2v_i k]^2}-\frac{1}{(v_s-v_i)^2 q^2}\right)}.\label{SM:Eq:GF_IR}
\end{align}

\subsection{The zero-momentum Green's function $|G(k=0,t;k=0,0)|$} \label{SM:sec:GF_0}

In the zero-momentum case, we are able to derive a more precise form of the momentum-space Green's function. We start from the full expression of $|G(k,t;k,0)|$ specialized to the $k=0$ case and rewrite it as follows:
\begin{align}
\begin{split}
& |G(k=0,t;k=0,0)|= \Biggl|\exp{-\int du R(u;k=0)\frac{1-iut-e^{-iut}}{u^2}}\Biggr|=\exp{-\frac{g^2 K}{(2\pi)^2}\int_{-\infty}^{+\infty}dp\, |p|e^{-\alpha|p|}\frac{1-\cos\left[(v_i+v_s)|p|t\right]}{(v_i+v_s)^2|p|^2}}=  \\
 &= \exp{-\frac{g^2 K}{(2\pi)^2}2\int_{0}^{+\infty}dp\, e^{-\alpha p}\frac{1-\cos\left[(v_i+v_s)pt\right]}{(v_i+v_s)^2 p}}=\exp{-\frac{g^2 K}{(2\pi)^2}\frac{2}{(v_i+v_s)^2}\int_0^{+\infty} dx\, e^{-\frac{\alpha x}{(v_i+v_s)t}}\frac{1-\cos x}{x}},
 \end{split}
\end{align}
where, in the last step, we performed the substitution $x=(v_i+v_s)pt$. Thus, the large-time asymptotic behavior of $|G(k=0,t;k=0,0)|$ depends on the large-time behavior of the integral expression $I(t)=\int_0^{+\infty} dx\, e^{-\frac{\alpha x}{(v_i+v_s)t}}\frac{1-\cos x}{x}$. We evaluate the latter first by subdividing the range of integration as:
\begin{align}
\begin{split}
&I(t)=\\
&=\int_0^{1} dx\, e^{-\frac{\alpha x}{(v_i+v_s)t}}\frac{1-\cos x}{x}+\int_1^{\frac{(v_i+v_s)t}{\alpha}} dx\, e^{-\frac{\alpha x}{(v_i+v_s)t}}\frac{1-\cos x}{x}+\int_{\frac{(v_i+v_s)t}{\alpha}}^{\frac{(v_i+v_s)^2 t^2}{\alpha^2}} dx\, e^{-\frac{\alpha x}{(v_i+v_s)t}}\frac{1-\cos x}{x}+\\
&+\int_{\frac{(v_i+v_s)^2 t^2}{\alpha^2}}^{+\infty} dx\, e^{-\frac{\alpha x}{(v_i+v_s)t}}\frac{1-\cos x}{x}=I_1(t)+I_2(t)+I_3(t)+I_4(t),
\end{split}
\end{align}
and then considering the role of each contribution when $t\gg 1$.

In such limit, it is easy to see that $I_1(t)$ is bounded by the time-independent $O(1)$ constant $\int_0^1 dx \frac{1-\cos x}{x}$. Similarly, exploiting the fact that $\frac{1-\cos x}{x}\leq \frac{2}{x}$ for $x>0$, one obtains the following estimate on the value of $I_4(t)$:
\begin{align}
I_4(t)=\int_{\frac{(v_i+v_s)^2 t^2}{\alpha^2}}^{+\infty}e^{-\frac{\alpha x}{(v_i+v_s)t}}\frac{1-\cos x}{x}\leq \frac{2\alpha^2}{(v_i+v_s)^2 t^2}\int _{\frac{(v_i+v_s)^2 t^2}{\alpha^2}}^{+\infty}e^{-\frac{\alpha x}{(v_i+v_s)t}}= \frac{2\alpha}{(v_i+v_s)t}e^{-\frac{(v_i+v_s)t}{\alpha}}\xrightarrow{t\rightarrow +\infty} 0.
\end{align}
Similar considerations hold for $I_3(t)$, which turns out to be bounded by a $O(1)$ constant in the large-time limit:
\begin{align}
    I_3(t)=\int_{\frac{(v_i+v_s)t}{\alpha}}^{\frac{(v_i+v_s)^2 t^2}{\alpha^2}} dx\, e^{-\frac{\alpha x}{(v_i+v_s)t}}\frac{1-\cos x}{x} \leq \frac{2\alpha}{(v_i+v_s)t}\int_{\frac{(v_i+v_s)t}{\alpha}}^{\frac{(v_i+v_s)^2 t^2}{\alpha^2}} dx\, e^{-\frac{\alpha x}{(v_i+v_s)t}}=2\left(e^{-1}-e^{-\frac{(v_i+v_s)t}{\alpha}} \right)\xrightarrow{t\rightarrow +\infty} 2e^{-1}.
\end{align}
Finally, we estimate the most crucial term, namely $I_2(t)$, by firstly noticing that the cosin term only contributes a finite $O(1)$ constant in the large-time limit, so that:
\begin{align}
    I_2(t)\sim\int_1^{\frac{(v_i+v_s) t}{\alpha}} dx\, e^{-\frac{\alpha x}{(v_i+v_s)t}}\frac{1}{x}.
\end{align}
After an integration by parts, the latter expression is rewritten as:
\begin{align}
\begin{split}
& e^{-1}\log\left(\frac{v_i+v_s}{\alpha}t\right)+\frac{\alpha}{(v_i+v_s)t}\int_1^{\frac{(v_i+v_s)t}{\alpha}} dx\, e^{-\frac{\alpha x}{(v_i+v_s)t}} \log x=\\
&=e^{-1}\log\left(\frac{v_i+v_s}{\alpha}t\right)+\log\left(\frac{v_i+v_s}{\alpha}t\right)\left(e^{-\frac{\alpha}{(v_i+v_s)t}}-e^{-1} \right)+\int_{\frac{\alpha}{(v_i+v_s)t}}^1 dy \,e^{-y} \log y,\label{SM:Eq:int_GF_k}
\end{split}
\end{align}
where in the last step we performed the substitution $y=\frac{\alpha x}{(v_i+v_s)t}$. Since the last term in ~\eqref{SM:Eq:int_GF_k} converges to a constant $O(1)$ value for large values of $t$, one obtains in the end that:
\begin{align}
    I_2(t)
\sim^{t\gg 1}\log\left(\frac{v_i+v_s}{\alpha}t\right),
\end{align}
thus recovering the result in~\eqref{Eq:GF_0}.

\subsection{Effect of the interaction strength $g$ on $|G(x,t;0,0)|$}

\begin{figure}
\includegraphics[width=0.48\textwidth]{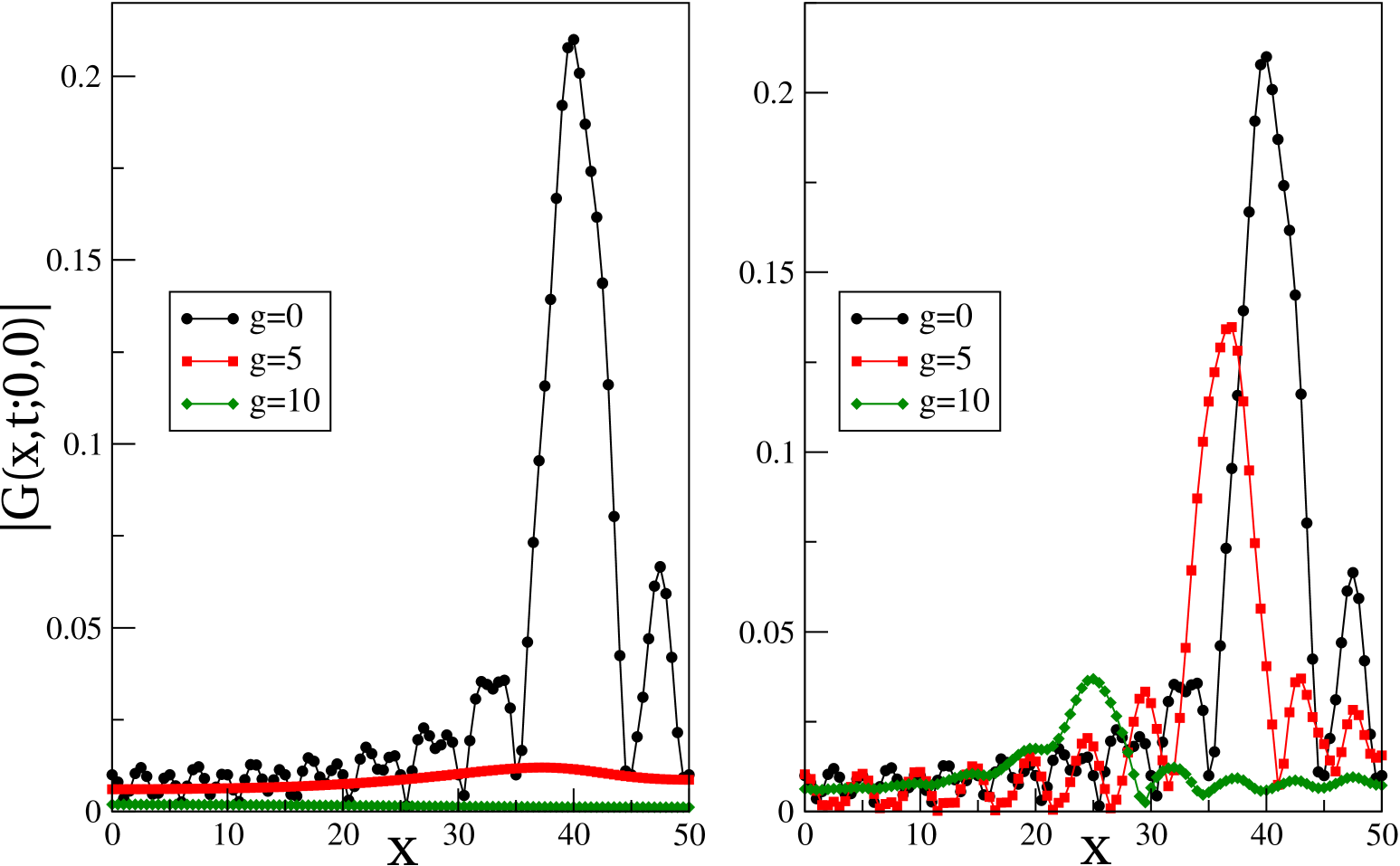}
\caption{\label{Fig:GF_g} Left: $|G(x,t;0,0)|$ as a function of the interaction strength $g$ in the case $v_i>v_s$
for parameter values $L=100,\, v_s=1,\, v_i=4,\, K=1.5,\, \alpha=10^{-1}$.Right: $|G(x,t;0,0)|$ as a function of the interaction strength $g$ in the case $v_i<v_s$
for $v_s=7$, the other parameters being unchanged.}
\label{SM:Fig:GF_g}
\end{figure}

As discussed in the main text, we provide in Fig.~\ref{SM:Fig:GF_g} the effect of an increasing interaction strength $g$ on the propagation speed and coherence of the impurity. The interactions both slow down the impurity propagation and destroy its quasiparticle coherence. Moreover, the effect is much more dramatic in the case $v_i>v_s$, as expected.

\section{The impurity density profile}\label{SM:density}

Considering an initial state for the impurity of the form:
\begin{equation}
\ket{\psi(0)}= \int_{-\frac{L}{2}}^{\frac{L}{2}}dy\,\, \psi(y) \hat d^{\dag}(y)\ket{0}.
\end{equation}
Then, the expectation value of the density in the time-evolved state $\ket{\psi(t)}=e^{-i\hat H t}\ket{\psi(0)}$ can be rewritten in the following way:
\begin{align}\label{Eq:density_ev}
\bra{\psi(t)}\hat d^{\dag}(x) \hat d(x)\ket{\psi(t)}=\frac{1}{L^2}\sum_{k,k', q,q'} e^{-i(k-k')x}\int_{-\frac{L}{2}}^{\frac{L}{2}} dy\, \psi^*(y) e^{iqy} \int_{-\frac{L}{2}}^{\frac{L}{2}} dz\, \psi(z) e^{-iq'z} \bra{0}\hat d_q e^{i\hat H t}\hat d^{\dag}_k \hat d_{k'}e^{-i\hat H t}\hat d^{\dag}_{q'} \ket{0}.
\end{align}
Hence, the problem of computing $\bra{\psi(t)}\hat \rho(x) \ket{\psi(t)}$ boils down to the calculation of the matrix element in~\eqref{Eq:density_ev}.

We rewrite the matrix element in~\eqref{Eq:density_ev} for the impurity density profile as a functional integral, namely:
\begin{align}
\begin{split}
\bra{0}\hat d_{k_1} e^{i\hat H t}\hat d^{\dag}_{k_2} \hat d_{k_3}e^{-i\hat H t}\hat d^{\dag}_{k_4} \ket{0}&= \int \mathcal{D}\phi \mathcal{D}\varphi e^{i(S_0+S_{int})} \varphi_{k_1,-}(0)\varphi_{k_2,-}^*(t)\varphi_{k_3,+}(t)\varphi_{k_4,+}^*(0)=\\
&=\langle  \varphi_{k_1,-}(0)\varphi_{k_2,-}^*(t)\varphi_{k_3,+}(t)\varphi_{k_4,+}^*(0) e^{iS_{int}}  \rangle_{S_0}.
\end{split}
\end{align}

\subsection{Zeroth-order term}

The first term in the perturbative expansion corresponds to the noninteracting average:
\begin{align}
\langle  \varphi_{k_1,-}(0)\varphi_{k_2,-}^*(t)\varphi_{k_3,+}(t)\varphi_{k_4,+}^*(0) \rangle_{S_0}=\delta_{k_1,k_2}\delta_{k_3,k_4}e^{i(\epsilon_{k_1}-\epsilon_{k_3})t},
\end{align}
as one can show both by means of the Schroedinger equation and of the Keldysh path-integral formulation. In following the latter route, one needs to exploit the expressions in~\eqref{SM:Eq:corr1}\eqref{SM:Eq:corr2}\eqref{SM:Eq:corr3}\eqref{SM:Eq:corr4},
where we remind that $F(\epsilon)=\coth\left[\beta(\epsilon-\mu)/2\right]=1+2n_B(\epsilon)$ for a noninteracting system at equilibrium with inverse temperature $\beta$ and chemical potential $\mu$. Since the initial density matrix of the bosonic impurity is $\ket{0}\bra{0}$, namely the state with zero occupancy for all the modes, we set $n_B(\epsilon)=0$ in the expression of $F(\epsilon)$ in the case of the impurity degrees of freedom.

\subsection{Second-order term}

Next, one needs to evaluate:
\begin{align}
-\langle  \varphi_{k_1,-}(0)\varphi_{k_2,-}^*(t)\varphi_{k_3,+}(t)\varphi_{k_4,+}^*(0)\frac{1}{2}S_{int}^2 \rangle_{S_0}. \label{SM:Eq:sec_ord}
\end{align}
Plugging in the expression of $S_{int}$,
and, for the sake of convenience, introducing the notation:
\begin{align}
& G(k,\alpha,t|k',\beta,t'):=\langle \varphi_{k,\alpha}(t)\varphi^*_{k',\beta}(t') \rangle,\\
&G_{ph}(k,\alpha,t|k',\beta,t'):=\langle \phi_{k,\alpha}(t)\phi^*_{k',\beta}(t') \rangle,
\end{align}
one obtains the following expression for the second-order contribution in~\eqref{SM:Eq:sec_ord}:
\begin{align}
\begin{split}
&-\frac{g^2 K}{2L^2}\sum_{k,k'}\sum_{p,p'\neq 0}V(p) V(p')\int dt_1\int dt_2 \\
&\Biggl\{ \Biggr[G(k_1,-,0|k_2,-,t)G(k_3,+,t|k,+,t_1)G(k+p,+,t_1|k',+,t_2)G(k'+p',+,t_2|k_4,+,0)+\\
&+G(k_1,-,0|k_2,-,t)G(k_3,+,t|k',+,t_2)G(k'+p',+,t_2|k,+,t_1)G(k+p,+,t_1|k_4,+,0)\Biggr]\times\\
&\times\Biggl[G_{ph}(-p',+,t_2|p,+,t_1)+G_{ph}(-p,+,t_1|p',+,t_2) \Biggr]+\\
&-G(k_1,-,0|k',-,t_2)G(k'+p',-,t_2|k_2,-,t)G(k_3,+,t|k,+,t_1)G(k+p,+,t_1|k_4,+,0)\times\\
&\times\Biggl[G_{ph}(-p',-,t_2|p,+,t_1)+G_{ph}(-p,+,t_1|p',-,t_2) \Biggr]+\\
&-G(k_1,-,0|k,-,t_1)G(k+p,-,t_1|k_2,-,t)G(k_3,+,t|k',+,t_2)G(k'+p',+,t_2|k_4,+,0)\times\\
&\times\Biggl[G_{ph}(-p',+,t_2|p,-,t_1)+G_{ph}(-p,-,t_1|p',+,t_2) \Biggr]+\\ 
&+\Biggr[G(k_1,-,0|k,-,t_1)G(k+p,-,t_1|k',-,t_2)G(k'+p',-,t_2|k_2,-,t)G(k_3,+,t|k_4,+,0)+\\
&+G(k_1,-,0|k',-,t_2)G(k'+p',-,t_2|k,-,t_1)G(k+p,-,t_1|k_2,-,t)G(k_3,+,t|k_4,+,0)\Biggr]\times\\
&\times\Biggl[G_{ph}(-p',-,t_2|p,-,t_1)+G_{ph}(-p,-,t_1|p',-,t_2) \Biggr]\Biggr\} \label{Eq:matrix_element}
\end{split}
\end{align}

\subsection{Final expression of the density within a second order expansion}

After plugging the result shown in~\eqref{Eq:matrix_element} into~\eqref{Eq:density_ev} and defining the Fourier transform of the initial wavefunction as:
\begin{equation}\label{Eq:FT}
(\mathcal{F}\psi)(q):=\frac{1}{L}\int_{-\frac{L}{2}}^{\frac{L}{2}} dx\, e^{-iqx}\psi(x),
\end{equation}
one obtains the final result:
\begin{align}
\begin{split}
&\bra{\psi(t)}\hat \rho(x) \ket{\psi(t)}=\sum_{k,k_2} e^{-i(k_2-k)x}e^{i(\epsilon_{k_2}-\epsilon_k)t}(\mathcal{F}\psi)^*(k_2)(\mathcal{F}\psi)(k)+\\
&-\frac{g^2 K}{2L^2}\sum_{k,k_2}\sum_{p\neq 0} V^{2}(p) e^{-i(k_2-k)x}(\mathcal{F}\psi)^*(k_2)(\mathcal{F}\psi)(k)\times\\
&\times \Bigg\{ e^{i(\epsilon_{k_2}-\epsilon_k)t}\Biggl [\frac{F(\omega_p)-1}{2}\int_{0}^t dt_2\int_{t_2}^t dt_1 e^{i(\epsilon_k+\omega_p-\epsilon_{k+p})(t_1-t_2)}+\frac{F(\omega_p)+1}{2}\int_{0}^t dt_2\int_{t_2}^t dt_1 e^{i(\epsilon_k-\epsilon_{k+p}-\omega_p)(t_1-t_2)} \Biggr]  +\\
&+ e^{i(\epsilon_{k_2}-\epsilon_k)t}\Biggl [\frac{F(\omega_p)+1}{2}\int_{0}^t dt_1\int_{t_1}^t dt_2 e^{i(\epsilon_{k+p}+\omega_p-\epsilon_{k})(t_1-t_2)}+\frac{F(\omega_p)-1}{2}\int_{0}^t dt_1\int_{t_1}^t dt_2 e^{i(\epsilon_{k+p}-\epsilon_{k}-\omega_p)(t_1-t_2)} \Biggr] +\\
& - e^{i(\epsilon_{k_2+p}-\epsilon_{k+p})t}\Biggl [\frac{F(\omega_p)+1}{2}\int_{0}^t dt_1\int_{0}^t dt_2 e^{i(\epsilon_{k_2}-\omega_p-\epsilon_{k_2+p})t_2}e^{i(\epsilon_{k+p}+\omega_p-\epsilon_k)t_1}+\\
&\qquad\qquad\qquad\qquad+\frac{F(\omega_p)-1}{2}\int_{0}^t dt_1\int_{0}^t dt_2 e^{i(\epsilon_{k_2}-\epsilon_{k_2+p}+\omega_p)t_2}e^{i(\epsilon_{k+p}-\epsilon_k-\omega_p)t_1} \Biggr]+\\
&- e^{i(\epsilon_{k_2+p}-\epsilon_{k+p})t}\Biggl [\frac{F(\omega_p)-1}{2}\int_{0}^t dt_1\int_{0}^t dt_2 e^{i(\epsilon_{k_2}+\omega_p-\epsilon_{k_2+p})t_1}e^{i(\epsilon_{k+p}-\epsilon_k-\omega_p)t_2}+\\
&\qquad\qquad\qquad\qquad+\frac{F(\omega_p)+1}{2}\int_{0}^t dt_1\int_{0}^t dt_2 e^{i(\epsilon_{k_2}-\epsilon_{k_2+p}-\omega_p)t_1}e^{i(\epsilon_{k+p}+\omega_p-\epsilon_k)t_2} \Biggr] +\\
&+ e^{i(\epsilon_{k_2}-\epsilon_k)t}\Biggl [\frac{F(\omega_p)-1}{2}\int_{0}^t dt_1\int_{t_1}^t dt_2 e^{i(\epsilon_{k_2}+\omega_p-\epsilon_{k_2+p})(t_1-t_2)}+\frac{F(\omega_p)+1}{2}\int_{0}^t dt_1\int_{t_1}^t dt_2 e^{i(\epsilon_{k_2}-\epsilon_{k_2+p}-\omega_p)(t_1-t_2)} \Biggr] +\\
&+ e^{i(\epsilon_{k_2}-\epsilon_k)t}\Biggl [\frac{F(\omega_p)+1}{2}\int_{0}^t dt_2\int_{t_2}^t dt_1 e^{i(\epsilon_{k_2+p}+\omega_p-\epsilon_{k_2})(t_1-t_2)}+\frac{F(\omega_p)-1}{2}\int_{0}^t dt_2\int_{t_2}^t dt_1 e^{i(\epsilon_{k_2+p}-\epsilon_{k_2}-\omega_p)(t_1-t_2)} \Biggr]\Biggr\}.
\end{split}
\end{align}

\subsubsection{$T=0$ limit}

In the limit where $T\rightarrow 0^+$, the phononic distribution function $F(\omega_p)=\coth(\beta\omega_p/2)$ satisfies:
\begin{equation}
\lim_{\beta\rightarrow +\infty} F(\omega_p)=1,
\end{equation}
since $\omega_p >0\,\,\forall p\neq 0$.
In such a case the expression for the time-evolved density profile simplifies to:
\begin{align}
\begin{split}
&\bra{\psi(t)}\hat \rho(x) \ket{\psi(t)}=\\
&=\sum_{k,k_2} e^{-i(k_2-k)x}e^{i(\epsilon_{k_2}-\epsilon_k)t}(\mathcal{F}\psi)^*(k_2)(\mathcal{F}\psi)(k)-\frac{g^2 K}{2L^2}\sum_{k,k_2}\sum_{p\neq 0} V^{2}(p) e^{-i(k_2-k)x}(\mathcal{F}\psi)^*(k_2)(\mathcal{F}\psi)(k)\times\\
&\times \Bigg\{ e^{i(\epsilon_{k_2}-\epsilon_k)t}\int_{0}^t dt_2\int_{t_2}^t dt_1 e^{i(\epsilon_k-\epsilon_{k+p}-\omega_p)(t_1-t_2)}  + e^{i(\epsilon_{k_2}-\epsilon_k)t}\int_{0}^t dt_1\int_{t_1}^t dt_2 e^{i(\epsilon_{k+p}+\omega_p-\epsilon_{k})(t_1-t_2)} +\\
& - e^{i(\epsilon_{k_2+p}-\epsilon_{k+p})t}\int_{0}^t dt_1\int_{0}^t dt_2 e^{i(\epsilon_{k_2}-\omega_p-\epsilon_{k_2+p})t_2}e^{i(\epsilon_{k+p}+\omega_p-\epsilon_k)t_1}+\\
&- e^{i(\epsilon_{k_2+p}-\epsilon_{k+p})t}\int_{0}^t dt_1\int_{0}^t dt_2 e^{i(\epsilon_{k_2}-\epsilon_{k_2+p}-\omega_p)t_1}e^{i(\epsilon_{k+p}+\omega_p-\epsilon_k)t_2}  +\\
&+ e^{i(\epsilon_{k_2}-\epsilon_k)t}\int_{0}^t dt_1\int_{t_1}^t dt_2 e^{i(\epsilon_{k_2}-\epsilon_{k_2+p}-\omega_p)(t_1-t_2)}  + e^{i(\epsilon_{k_2}-\epsilon_k)t}\int_{0}^t dt_2\int_{t_2}^t dt_1 e^{i(\epsilon_{k_2+p}+\omega_p-\epsilon_{k_2})(t_1-t_2)} \Biggr\}=\\
&=\sum_{k,k_2} e^{-i(k_2-k)x}e^{i(\epsilon_{k_2}-\epsilon_k)t}(\mathcal{F}\psi)^*(k_2)(\mathcal{F}\psi)(k)\Biggl\{ 1-\frac{g^2 K}{L^2}\sum_{p\neq 0} V^2 (p)\Biggl [\int du \delta(u-\epsilon_k+\epsilon_{k+p}+\omega_p) \frac{1+iut-e^{iut}}{u^2}+\\
&+\int du \delta(u-\epsilon_{k_2}+\epsilon_{k_2+p}+\omega_p) \frac{1-iut-e^{-iut}}{u^2}+\\
&-\int du\int dv \delta(u-\epsilon_{k_2}+\epsilon_{k_2+p}+\omega_p) \delta(v-\epsilon_{k+p}-\omega_p+\epsilon_{k}) \frac{1-e^{-iut}}{iu}\frac{1-e^{-ivt}}{iv}\Biggr]  \Biggr\}. \label{SM:Eq:density_calc}
\end{split}
\end{align}
Reexponentiating the term in curly brackets of the last expression in~\eqref{SM:Eq:density_calc} and introducing the functions:
\begin{align}
 &R(u;k)=\frac{g^2 K}{(2\pi)^2}\int dp |p| e^{-\alpha|p|}\delta(u-\epsilon_{k+p}-\omega_p+\epsilon_k),\label{Eq:R}\\
 & S(u,v;k,k_2)=\frac{g^2 K}{(2\pi)^2}\int dp |p| e^{-\alpha|p|}\delta(u-\epsilon_{k_2+p}-\omega_p+\epsilon_{k_2})\delta(v-\epsilon_{k+p}-\omega_p+\epsilon_k)
 \end{align}
one obtains the result presented in~\eqref{Eq:density_LCE}:
\begin{align}\label{Eq:density}
\begin{split}
&\bra{\psi(t)}\hat \rho(x) \ket{\psi(t)}= \sum_{k,k_2}(\mathcal{F}\psi)^*(k_2)(\mathcal{F}\psi)(k) e^{i[(kx-\epsilon_k t) - (k_2 x-\epsilon_{k_2}t)]}\times\\
& \times e^{-\left[\int du R(u;k)\frac{1-iut-e^{-iut}}{u^2} +\int du R(u;k_2)\frac{1+iut-e^{iut}}{u^2}-\int du\int dv S(u,v;k,k_2) \frac{1-e^{iut}}{u}\frac{1-e^{-ivt}}{v}\right]}
\end{split}
\end{align}

It is worth noticing that the functions appearing in the expression above possess a number of symmetry properties that constrain their explicit expression, namely:
\begin{align}
& g(v,u,t)=g^* (u,v,t), \label{Eq:g_prop}\\
& S(v,u;k_2,k)= S(u,v;k,k_2),\\
& S(u,v,-k,-k_2) = S(u,v; k,k_2),\\
& R(u;-k) = R(u;k),\\
& -\int du R(u;k)\left[f(u,t)+f^*(u,t) \right]+\int du dv S(u,v;k,k) g(u,v,t)=0. \label{Eq:exp_vanishing}
\end{align}

\end{document}